\documentclass[conference]{IEEEtran}
\IEEEoverridecommandlockouts
\usepackage[left=0.75in,right=0.75in,top=0.76in,bottom=1.2in]{geometry}
\usepackage{cite}
\usepackage{amsmath,amssymb,amsfonts}
\usepackage{algorithmic}
\usepackage{graphicx}
\usepackage{pdfpages}
\usepackage{textcomp}
\usepackage{xcolor}
\usepackage{adjustbox}
\usepackage{booktabs}
\usepackage{float} % for [H] float option
\usepackage[font=footnotesize,labelfont=bf]{caption}
\usepackage{dblfloatfix} % for placing figure at bottom
\usepackage{graphics} % for pdf, bitmapped graphics files
\usepackage{amsmath} % assumes amsmath package installed
\usepackage{amssymb}  % assumes amsmath package installed
\usepackage{cite}
\usepackage{xcolor}
\usepackage{amsmath,amssymb,amsfonts}
\usepackage{xspace}
\usepackage{multirow}
\usepackage{array}
\usepackage{url}
\usepackage{breakurl}
\usepackage{comment}
\newcommand{\hide}[1]{}
\usepackage[utf8]{inputenc} % for UTF-8 source files
\usepackage{textcomp}       % for some text symbols
\usepackage{tabularx} % in preamble
\usepackage{url}
\usepackage{enumitem}
\usepackage{listings}
\def\BibTeX{{\rm B\kern-.05em{\sc i\kern-.025em b}\kern-.08em
    T\kern-.1667em\lower.7ex\hbox{E}\kern-.125emX}}

\begin{document}

\title{Supply Chain Exploitation of Secure ROS~2 Systems: A Proof-of-Concept on Autonomous Platform Compromise via Keystore Exfiltration}

% \author{
%     \IEEEauthorblockN{
%         Tahmid Hasan Sakib*\IEEEauthorrefmark{1}\thanks{*Tahmid Hasan Sakib and Yago Romano Martinez contributed equally to this work.},
%         Yago Romano Martinez*\IEEEauthorrefmark{2},
%         Brady Carter\IEEEauthorrefmark{1},
%         Syed Rafay Hasan\IEEEauthorrefmark{1},
%         and Terry N. Guo\IEEEauthorrefmark{3}
%     }
%     \IEEEauthorblockA{\IEEEauthorrefmark{1}Department of Electrical and Computer Engineering, Tennessee Technological University, Cookeville, TN 38505, USA\\
%     Email: \{tsakib42, clbrady43, shasan\}@tntech.edu}
%     \IEEEauthorblockA{\IEEEauthorrefmark{2}Department of Computer Science, Tennessee Technological University, Cookeville, TN 38505, USA\\
%     Email: yromanoma42@tntech.edu}
%     \IEEEauthorblockA{\IEEEauthorrefmark{3}Center for Manufacturing Research, Tennessee Technological University, Cookeville, TN 38505, USA\\
%     Email: nguo@tntech.edu}
% }

% \author{
%     Tahmid Hasan Sakib$^{1}$\thanks{*Tahmid Hasan Sakib and Yago Romano Martinez contributed equally to this work.}, 
%     Yago Romano Martinez$^{2*}$,
%     Carter Brady$^{1}$,
%     Syed Rafay Hasan$^{1}$,
%     and Terry N. Guo$^{3}$\\
%     $^{1}$Department of Electrical and Computer Engineering, Tennessee Tech University, Cookeville, TN, USA\\
%     $^{2}$Department of Computer Science, Tennessee Tech University, Cookeville, TN, USA\\
%     $^{3}$Center for Manufacturing Research, Tennessee Tech University, Cookeville, TN, USA\\
%     Email: \{tsakib42, yromanoma42, clbrady43, shasan, nguo\}@tntech.edu
% }

\author{
\IEEEauthorblockN{Tahmid Hasan Sakib, Yago Romano Martinez, Carter Brady, Syed Rafay Hasan, Terry N. Guo}
\thanks{
* Tahmid Hasan Sakib and Yago Romano Martinez contributed equally to this work. 

T.H. Sakib, C. Brady, and S.R. Hasan are with the Department of Electrical and Computer Engineering, Tennessee Technological University, Cookeville, TN, USA (email: tsakib42, clbrady43, shasan\{@tntech.edu\}).  
Y.R. Martinez is with the Department of Computer Science, Tennessee Technological University, Cookeville, TN, USA (email: yromanoma42@tntech.edu).  
T.N. Guo is with the Center for Manufacturing Research, Tennessee Technological University, Cookeville, TN, USA (email: nguo@tntech.edu).
}
}

\maketitle

\begin{abstract}
This paper presents a proof-of-concept supply chain attack against the Secure ROS~2 (SROS~2) framework, demonstrated on a Quanser QCar2 autonomous vehicle platform. A Trojan infected Debian package modifies core ROS~2 security commands to exfiltrate newly generated keystore credentials via DNS in base64-encoded chunks to an attacker-controlled nameserver. Possession of these credentials enables the attacker to rejoin the SROS~2 network as an authenticated participant and publish spoofed control or perception messages without triggering authentication failures. We evaluate this capability on a secure ROS~2 Humble testbed configured for a four-stop-sign navigation routine using an Intel RealSense camera for perception. Experimental results show that control-topic injections can cause forced braking, sustained high-speed acceleration, and continuous turning loops, while perception-topic spoofing can induce phantom stop signs or suppress real detections. The attack generalizes to any data distribution service (DDS)-based robotic system using SROS~2, highlighting the need for both supply chain integrity controls and runtime semantic validation to safeguard autonomous systems against insider and impersonation threats.


\end{abstract}
% \vspace{-4mm}

\section{Introduction}
\label{sec:intro}
Cyber-physical systems (CPS) with increasing levels of autonomy are now deployed in industrial automation, transportation, defense, and research, operating in complex networked environments where failures can have safety-critical consequences \cite{m1_4519604,m2_7924372}. Autonomous vehicle (AV) security researchers consistently highlight that malicious interference with perception, planning, or control subsystems can lead to hazardous outcomes\cite{m3_6899663,m4_7935369,m5_checkoway2011comprehensive,m35_11030565}. The Robot Operating System (ROS) has emerged as the dominant middleware for robotics, enabling modular, distributed development across heterogeneous platforms\cite{m6_quigley2009ros, m7_macenski2022robot}. ROS~2, a complete re-architecture built on the Data Distribution Service (DDS) standard, offers real-time publish–subscribe communication with support for multi-vendor interoperability\cite{m10_Fazzari2019,m11_sros2_repo}. Its adoption spans both simulated environments such as CARLA\cite{m31_dosovitskiy2017carla} and real-world testbeds\cite{m30_liu2020novel}, underscoring the need for robust security across deployment contexts.

% Paragraph 2 – Known attack surfaces in ROS~2 systems
Early work on ROS1 showed that, without authentication or encryption, any host that could reach the ROS master could subscribe to sensitive topics or publish arbitrary commands~\cite{m8_garzon2017using,m9_dieber2017security}. SROS1 attempted to mitigate these issues, but adoption is limited. In ROS~2, DDS-Security is part of the middleware: when enabled, participants use certificates for identity, encrypted transport, and permissions defined by governance/permissions XML. SROS~2 provides the tools and policies to configure this stack~\cite{m10_Fazzari2019,m11_sros2_repo}.
However, even with these protections, critical security gaps have been discovered. Deng et al. identified four distinct vulnerabilities in SROS~2 that allow an attacker to bypass access controls, including unauthorized subscription and message injection, even when security is enabled\cite{m12_deng2022security}. Additionally, a known flaw in SROS~2’s default certificate handling permits nodes to self-sign and validate unauthorized permission files when using a single Certificate Authority (CA), effectively breaking the chain of trust and enabling identity spoofing\cite{m13_sros2_issue_282}. Beyond these specific exploits, the complexity of SROS~2 configuration and its performance overhead have discouraged adoption. Studies show that latency and throughput penalties introduced by DDS security plugins often lead developers to disable these protections in practice\cite{m14_kim2018security}. Surveys of ROS~2 usage further confirm that security is frequently deprioritized in favor of performance or convenience\cite{m15_portugal2024ros3}. Misconfiguration remains widespread, especially with respect to DDS policy files and manual certificate management, as shown in recent formal vulnerability analyses\cite{m16_yang2024formal}. Simulation-based studies using CARLA and similar environments have also shown how spoofed control or perception nodes can mislead autonomous agents into unsafe actions, underscoring the need for robust trust mechanisms across both middleware and higher-level applications\cite{m17_app15137493}.

\begin{figure*}[htbp]
    \centering
     \includegraphics[width=0.93\textwidth]{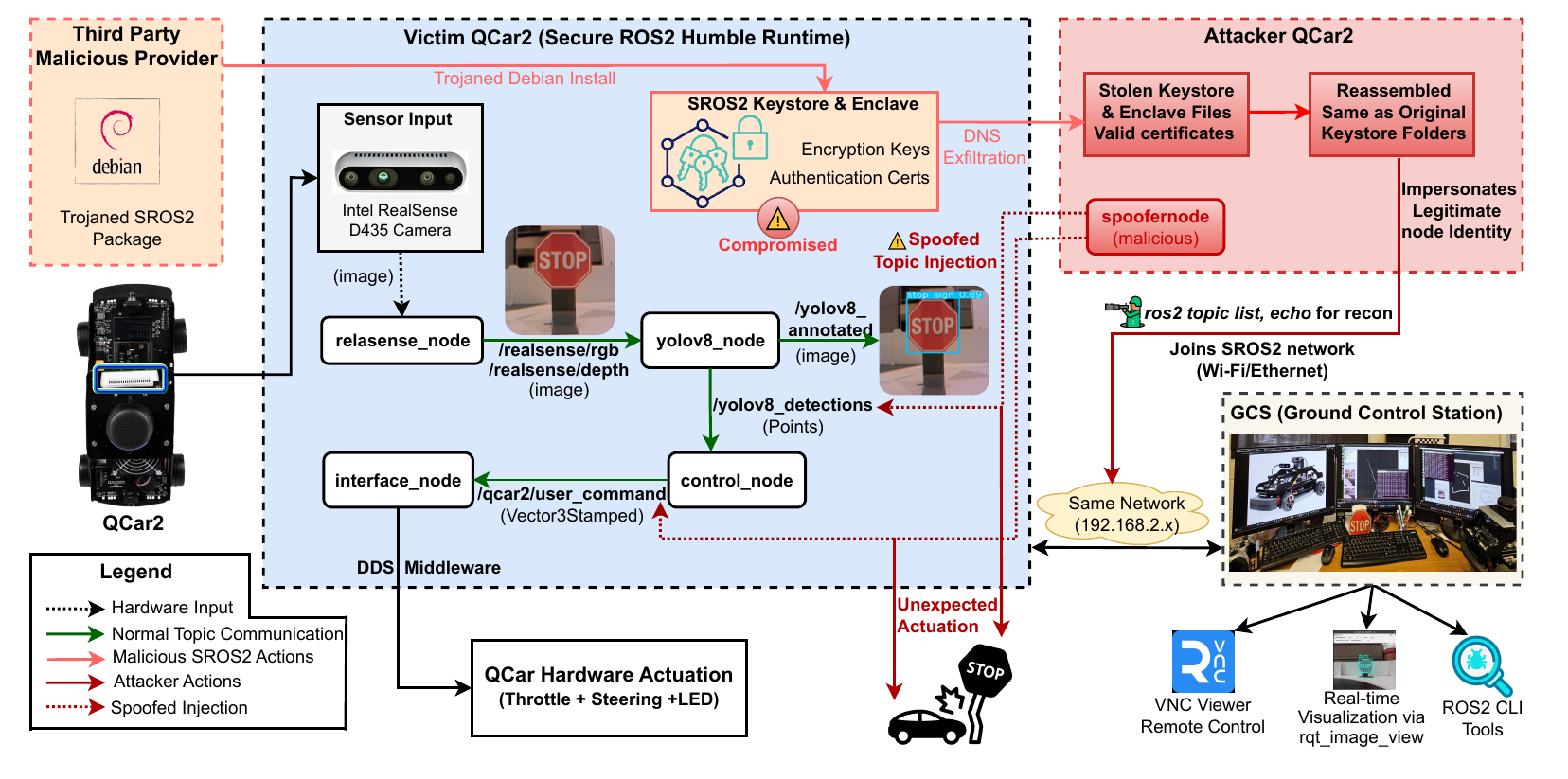}
    \caption{End-to-end supply-chain attack in a secure ROS~2 testbed: a malicious third-party package triggers DNS-based exfiltration of keystore credentials, enabling the attacker to rejoin the network and impersonate control or perception nodes.}
    \label{fig:SROS2_system_overview}
    \vspace{-5mm}
\end{figure*}

% Paragraph 3 – Motivation: SROS2 vulnerabilities via supply chain vectors
While much of the ROS~2 security literature focuses on direct network exploitation, the software supply chain represents a subtler but equally dangerous entry point\cite{m18_jang2014survey}. In 2019, Open Robotics reported a compromise of the official ROS build farm via a Jenkins vulnerability, prompting an emergency GPG key rotation to restore trust in distributed packages\cite{m19_kyrofa2019security,m20_ros_buildfarm2025}. Broader examples, such as the SolarWinds compromise\cite{m33_cisa2020_sunburst} and the 2024 XZ Utils backdoor attack (CVE-2024-3094)\cite{m21_nist2024_cve2024_3094,m22_claburn2024_xz_backdoor}, demonstrate how malicious logic can be inserted into trusted software without immediate detection. Once deployed, precompiled packages, which are common in ROS ecosystems, are difficult to audit for malicious code. Prior malware campaigns hosted on repositories such as PyPI have shown that attackers use Domain Name System (DNS) queries as a covert exfiltration channel for sensitive files\cite{m23_pypi2024_domain_abuse,m24_toulas2022_pypi_credentials}. This is particularly relevant for SROS~2 because keystores and permission/governance XML files act as the root of trust for secure DDS communication \cite{m13_sros2_issue_282,m25_rti2025_access_control}. DNS traffic is often overlooked by intrusion detection systems, and techniques such as base64 encoding and small-chunk query splitting make such exfiltration difficult to detect\cite{m26_cova2010detection,m27_yadav2010detecting}. To the best of our knowledge, this is the first study to investigate supply-chain attacks in the context of SROS~2. While the risk of malicious ROS~2 packages has been acknowledged in general threat models\cite{m12_deng2022security}, existing literature has not examined how such packages could directly target and compromise SROS~2’s trust anchors.

% Paragraph 4 – Proof-of-concept overview
This work presents a proof-of-concept supply chain attack targeting SROS~2 credentials through a malicious third-party ROS~2 package. Unlike prior ROS~2 or DDS attack studies, which have focused on network level spoofing, runtime node compromise, or general supply chain risks, our approach directly exploits SROS~2’s credential generation process to obtain the keystore and XML policies before they are deployed. The attacker distributes a Debian (.deb) package containing a modified utility file that hooks into \texttt{ros2 security create\_keystore} and \texttt{ros2 security create\_enclave} operations\cite{m10_Fazzari2019,m11_sros2_repo}. When invoked, the payload automatically reads newly generated certificates, keys, and XML policy files, encodes them in base64, and transmits them in 32-byte chunks via DNS queries to an attacker-controlled server. The attacker then reconstructs the keystore and enclave structure offline, enabling them to rejoin the secure ROS~2 network as an authenticated participant. With this capability, the attacker can deploy spoofed nodes to publish falsified actuation commands or fabricated perception data. While our live demonstration uses a Quanser QCar2\cite{m34_quanser_qcar2} executing a four-stop-sign autonomous navigation routine, the technique is generalizable to any DDS-based robotic system relying on SROS~2.

% \vspace{-2mm}
% Paragraph 5 – Research questions
This study is guided by two research questions:
% \vspace{-5mm}
\begin{itemize}[leftmargin=*,nosep]
    \item \textbf{RQ1:} What is a feasible \emph{supply-chain attack model} against ROS~2/SROS~2, and can it be realized via a malicious third-party package that targets the SROS~2 CLI during keystore creation? 
    \item \textbf{RQ2:} Once credentials are stolen and network access is re-established, what end-to-end operational impacts may occur on an autonomous platform?
\end{itemize}

Our main contribution is a practical supply-chain attack model for ROS~2/SROS~2 with a reference implementation that exfiltrates keystores via DNS and demonstrates end-to-end impact on an autonomous platform. The remainder of this paper is organized as follows. Section II describes the SROS~2 Humble testbed. Section III presents the attack execution and its operational effects on the QCar2 platform. Section IV discusses potential mitigations, and Section V concludes the paper.

% \begin{figure*}[htbp]
%     \centering
%     \includegraphics[width=\textwidth,trim=0.32cm 0.59cm 0.32cm 0.2cm,clip]{SROS2_QCAR2.pdf}
%     \caption{End-to-end supply-chain attack in a secure ROS~2 testbed: a malicious third-party package triggers DNS-based exfiltration of keystore credentials, enabling the attacker to rejoin the network and impersonate control or perception nodes.}
%     \label{fig:SROS2_system_overview}
%     %\vspace{-5mm}
% \end{figure*}

%\vspace{-1.5mm}
% \section{ROS2 HUMBLE TESTBED IMPLEMENTATION ON QUANSER QCAR2 PLATFORM}
\section{SROS~2 Testbed and Threat Model Implementation on Quanser QCar2}
\vspace{-0.5mm}
This section presents our end-to-end setup and attack. Figure~\ref{fig:SROS2_system_overview} provides an overview of the full system and flow. We first state the environment and attacker position (System \& Threat Model), then describe the ROS~2/SROS~2 runtime on the QCar2 platform (Secure Runtime Setup). Next, we explain how the malicious \texttt{sros2} package is built and shipped (Trojan Package Creation), how newly created credentials are sent as DNS queries to a fixed-IP listener (DNS Exfiltration Mechanism), and how the recovered keystore is used to join the graph and publish as an authenticated node (Reassembly and Spoofed Node).
\vspace{-2.5mm}

\subsection{System \& Threat Model}
\vspace{-1mm}
We evaluate a secure autonomous vehicle (AV) testbed where a Quanser QCar2 and its control station communicate over a private LAN using ROS~2 Humble with SROS~2-enabled DDS Security. The attacker has no physical or root access and the vehicle has no general Internet access. Instead, the adversary operates upstream in the software supply chain as a malicious third-party distributor of SROS~2 packages. A tampered Debian package replaces the official ROS~2 security utilities with modified versions that silently exfiltrate keystore and enclave materials over DNS. The attacker controls an authoritative DNS namespace (e.g., \texttt{*.attacker.example}) and runs a DNS listener on the same private LAN at a fixed IP that directly receives UDP/53 queries. Figure~\ref{fig:SROS2_system_overview} summarizes the system architecture and end-to-end attack chain—from credential generation to DNS exfiltration and subsequent node impersonation on the secure DDS graph.
\vspace{1mm}

\noindent\textbf{Attacker capabilities: }Once the malicious package is installed and \texttt{ros2 security create\_keystore} or \texttt{create\_enclave} runs, the modified CLI (executing with the caller’s privileges):
\begin{enumerate}[leftmargin=*,nosep]
  \item reads the newly created keys, certificates, and permission/governance XMLs;
  \item sends the data as DNS-formatted queries to an attacker listener on the same LAN at a fixed IP (e.g., \texttt{10.0.0.53:53});
  \item reassembles the keystore on the attacker side and uses it to join the ROS~2 graph as an authenticated participant.
\end{enumerate}

\vspace{1mm}
\noindent\textbf{Why spoofing still works under SROS~2: }Because SROS~2 trust is identity-based, a node launched with stolen credentials is accepted by other nodes as legitimate. ROS~2 permits multiple publishers per topic, so an attacker can race or override \texttt{/qcar2/user\_command} and inject or suppress messages on \texttt{/detections} once admitted to the graph.

\vspace{-2mm}
\subsection{Secure Runtime Setup}
\vspace{-1mm}
Our testbed runs ROS~2 Humble (Ubuntu 22.04/Jammy) with SROS~2 enforcing DDS-Security between nodes:
\begin{itemize}[leftmargin=*,nosep]
    \item \textbf{camera\_node:} Publishes RealSense RGB/depth on \texttt{/rgb}.
    \item \textbf{yolov8\_node:} Subscribes \texttt{/rgb}, publishes detections on \texttt{/detections} and annotated frames on \texttt{/annotated}.
    \item \textbf{control\_node:} Consumes \texttt{/detections}, outputs actuation to \texttt{/qcar2/user\_command}.
    \item \textbf{interface\_node:} Applies \texttt{/qcar2/user\_command} to QCar2 motor/LED drivers.
\end{itemize}
A package-level SROS~2 keystore is used; each node runs in its own enclave. Credentials are created with \texttt{ros2 security create\_keystore} and \texttt{create\_enclave}, and enforced via \texttt{ROS\_SECURITY\_ROOT\_DIRECTORY} and \texttt{ROS\_SECURITY\_ENABLE}. For validation, \texttt{rqt\_image\_viewer} subscribes to \texttt{/annotated}. The pipeline operates on a private LAN without general Internet egress.%
\footnote{We also produced a Focal/20.04 build for Foxy-only environments; artifacts are OS/Python specific.}

\vspace{-2mm}
\subsection{Trojan Package Creation}
\begin{figure}[t]
  \centering
  \includegraphics[width=0.75\linewidth]{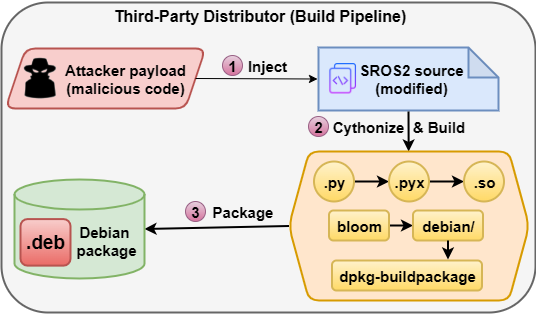}
  \caption{Build pipeline for a malicious SROS~2 Debian package at a third-party distributor.}
  \label{fig:trojan-build-pipeline}
  \vspace{-7mm}
\end{figure}

\vspace{-0.5mm}
We implemented a trojaned \texttt{sros2} CLI that exfiltrates credentials during keystore/enclave creation. The core payload resides in a modified \texttt{\_utilities.py} module, adding \texttt{exfil\_dns()} to read PKI artifacts (e.g., \texttt{.pem}, \texttt{.key}, \texttt{.xml}), base64-encode and chunk them, then transmit via DNS queries carrying file identifiers and chunk indices in the label. As illustrated in Figure~\ref{fig:trojan-build-pipeline}, selected \texttt{sros2} modules are converted to Cython (\texttt{.pyx}) and built into CPython extensions (\texttt{.so}); minimal \texttt{.py} stubs preserve imports while logic resides in compiled objects. Packaging used standard ROS~2 tooling: \texttt{bloom-generate rosdebian} to create \texttt{debian/} and \texttt{dpkg-buildpackage} via \texttt{pybuild} to produce a \texttt{.deb}. Post-install, the CLI interface is unchanged while the embedded payload runs with the invoking user’s privileges during normal \texttt{ros2 security} commands.

\vspace{-2mm}
\subsection{DNS Exfiltration Mechanism}
\vspace{-1mm}
Our payload uses DNS as a covert transport to leak SROS~2 credentials without spawning extra processes or requiring elevated privileges. After the malicious \texttt{sros2} package is installed, running \texttt{ros2 security create\_keystore} or \texttt{create\_enclave} triggers in-process enumeration and in-memory reads of the freshly created artifacts (e.g., \texttt{*.pem}, \texttt{*.key}, \texttt{permissions.xml}, \texttt{governance.xml}). Each file is Base64-encoded and split into fixed-size chunks (e.g., 32 bytes per chunk) to respect DNS label limits (~$\leq~63$~characters~per~label). Chunks are embedded in queries of the form:

\begin{lstlisting}
<filetag>.<chunk_data>.<chunk_index>.<session>.attacker.example
Example: 
perm.Y2hhbmtfYmxvY2tfMDAxPw==.001.a1.attacker.example
\end{lstlisting}
\vspace{-0.5mm}
The client sends DNS-formatted UDP packets to an attacker-run listener at a fixed LAN address (e.g., \texttt{10.0.0.53:53}). A short \texttt{<session>} token groups multi-file uploads. All logic runs within the CLI code path.

\vspace{-1.5mm}
\subsection{Reassembly and Spoofed Node Deployment}
\vspace{-0.5mm}
On the attacker side, the LAN DNS listener logs QNAMEs, extracts \texttt{<filetag>}, \texttt{<chunk\_data>}, \texttt{<chunk\_index>}, and \texttt{<session>}, then reorders and concatenates chunks before decoding to reconstruct the original keystore tree exactly (certificates, private keys, and permissions/governance XMLs). Possession of the keystore lets the attacker authenticate as a legitimate DDS participant. Because SROS~2 trust is identity-based, a node launched with stolen credentials is treated as a legitimate participant by other nodes. Spoofing modes are:
\begin{enumerate}[leftmargin=*,nosep]
    \item \textbf{Control Override} -- The attacker runs a fake publisher on \texttt{/qcar2/user\_command} (or a similar \texttt{/actuator\_command}). By sending messages more often, they can override normal commands and make the vehicle brake, accelerate, or steer.
    \item \textbf{Perception Injection} -- The attacker publishes fake sensor messages on \texttt{/detections} (e.g., false obstacles or hidden signs). These messages can mislead the planner into stopping, turning, or missing real hazards.
\end{enumerate}

In our QCar2 demonstration (Figure~\ref{fig:SROS2_system_overview}), the spoofed node altered the four-stop-sign routine (unexpected stops and turn changes), illustrating how credential compromise alone enables behavior manipulation in SROS~2 systems.

\begin{figure}[htbp]
    \centering
    \includegraphics[width=0.75\linewidth]{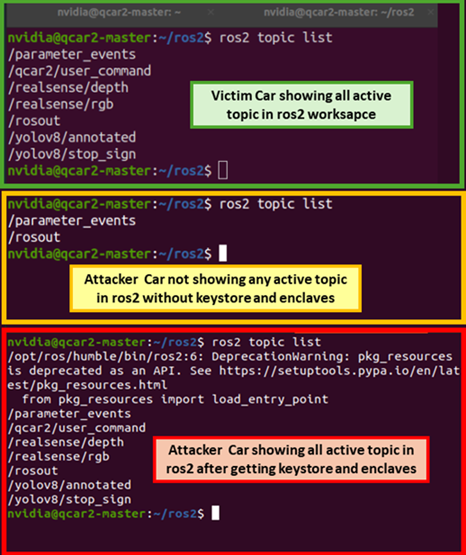}
    \caption{Topic discovery in a secured ROS~2 network: attacker can list topics only after stealing victim’s keystore and enclaves.}
    \label{fig:normalvsattacked}
    %\vspace{-5mm}
\end{figure}

\begin{figure}[htbp]
    \centering
    \includegraphics[width=0.8\linewidth]{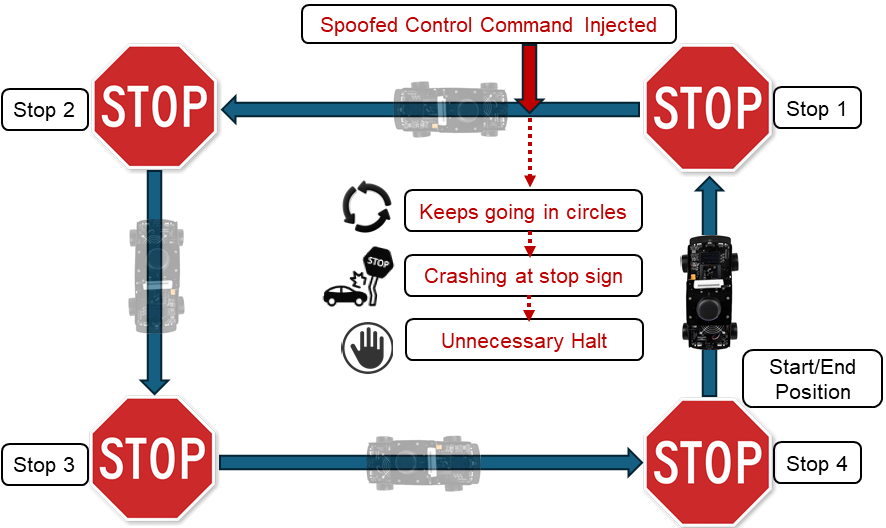}
    \caption{Navigation Routine with Spoofing Injection Points.}
    \label{fig:planned_path}
    \vspace{-6mm}
\end{figure}

%\vspace*{0.05in} % small adjustment
\begin{figure*}[!t]
  \centering
  \includegraphics[width=0.8\textwidth]{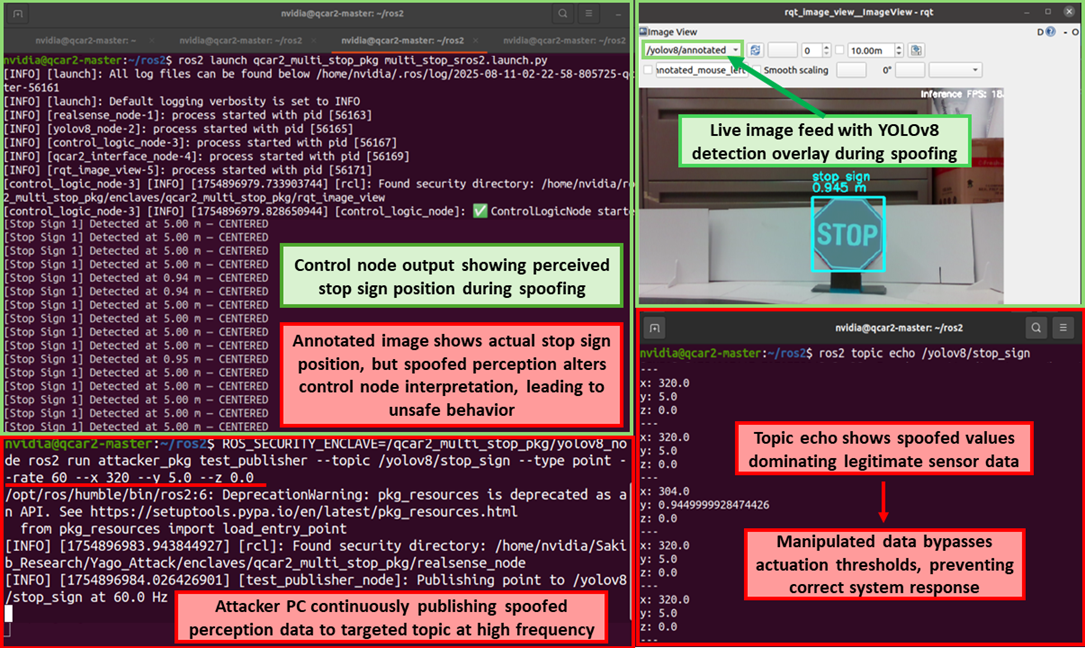}
  \caption{Representative spoofing attack on QCar2 ROS~2 topics, showing attacker injection of falsified data at high frequency, control node misinterpretation, and altered perception in the YOLOv8-annotated camera feed, leading to unsafe navigation outcomes.}
  \label{fig:spoofed-high-dist}
  \vspace{-2mm}
\end{figure*}

\begin{table*}[htbp]
\caption{Effects of Topic Spoofing on QCar2 Autonomous Navigation}
\label{tab:topic-spoofing}
\centering
\renewcommand{\arraystretch}{1.2}
\begin{tabularx}{\textwidth}{|l|X|X|X|}
\hline
\textbf{Topic Spoofed} & \textbf{Spoof Type} & \textbf{System Behavior} & \textbf{Outcome} \\ \hline
\verb|/qcar2/user_command| & Fixed throttle ($x=0.0$) & Override controlnode commands & Forced braking; halted despite clear path \\ \hline
\verb|/qcar2/user_command| & Fixed throttle ($x=1.0$) & Override controlnode commands & Sudden acceleration; high-speed crash \\ \hline
\verb|/yolov8/stop_sign| & Fixed steering ($x=\pm 0.6$) with mode change ($z=3$) & Override steering control & Abrupt turns; path deviation \\ \hline
\verb|/yolov8/stop_sign| & Low distance spoof ($y=0.6$\,m) & Triggered brake logic before stop sign & Premature stopping; unnecessary halt \\ \hline
\verb|/yolov8/stop_sign| & High distance spoof ($y=5.0$\,m) & Suppressed brake logic near stop sign & Failed to stop; collision risk \\ \hline
\end{tabularx}
\vspace{-4mm}
\end{table*}

Figure \ref{fig:normalvsattacked} illustrates the critical prerequisite for any spoofing attempt in our secured ROS~2 environment. In the top panel, the victim QCar2 lists all active topics in its ROS~2 workspace, including those for perception (\texttt{/yolov8/annotated}, \texttt{/yolov8/stop\_sign}), control (\texttt{/qcar/user\_command}), and raw sensor feeds. The middle panel in Figure \ref{fig:normalvsattacked} shows the attacker vehicle on the same network but without access to the keystore and enclave files; here, no protected topics are visible, as SROS~2 enforces access control at the DDS layer. Only after exfiltrating and reconstructing the victim’s keystore and enclave directories (bottom panel in Figure \ref{fig:normalvsattacked}) can the attacker’s node join the secure ROS~2 graph, enumerate all active topics, and inject or subscribe to messages indistinguishably from legitimate participants. This credential theft is therefore the enabler for both control-level and perception-level spoofing attacks detailed in Section \ref{sec:attack}.

\vspace{-1mm}
\section{Attack Execution and Effects}
\label{sec:attack}
\vspace{-0.5mm}
The spoofing attacks are carried out on the QCar2 while it executed the baseline navigation routine shown in Figure \ref{fig:planned_path}. A single representative example is shown in Figure \ref{fig:spoofed-high-dist}, which generalizes the behavior observed across multiple spoofed topics. In this example, the attacker injects falsified perception data at a high frequency, overriding legitimate sensor outputs in the control node’s decision-making loop. The manipulated values dominate the topic stream, altering control logic thresholds and triggering unsafe behavior of attackers choice. To illustrate the effectiveness of the attack we manifested few attacks that includes missing detection of stop signs, unnecessary braking, and erratic steering control.
\vspace{-3.5mm}

Across all spoofing trials, the attacker PC used an enclave-matching ROS~2 node to publish forged data at 60 Hz to a targeted topic. Although the visual example in Figure \ref{fig:spoofed-high-dist} shows a perception spoofing case, the same methodology applies to control command injection and LiDAR scan manipulation. In this example, legitimate stop sign detection data from the YOLOv8 node is overridden by spoofed values indicating a fixed, distant object location (5.0 m), despite the real stop sign being much closer. The control node, relying on this falsified perception, fails to trigger the stop command, demonstrating how manipulated data can bypass actuation thresholds and induce unsafe behavior. Each spoofing variant is tested in isolation to evaluate its distinct impact on autonomous navigation.

\enlargethispage{\baselineskip}
Table \ref{tab:topic-spoofing} summarizes the evaluated spoofing types, the targeted topics, the system behavior they induced, and the observed outcomes. Each row corresponds to one isolated spoofing trial, linking the injected values to their navigation impact. The selected topics (\texttt{/qcar/user\_command}, \texttt{/qcar/yolo\_detections}, \texttt{/qcar/scan}) are chosen because they span the primary perception-decision-actuation chain, meaning that that spoofing them can directly induce unsafe navigation behavior. The Spoof Type column describes the nature of the falsified data, chosen to simulate realistic adversary objectives such as forcing emergency stops or bypassing hazard detection. The Spoofed Values column records the specific forged parameters published during the trial, providing context for interpreting the downstream effects. The System Behavior column captures the real-time effect on the control node’s logic as observed during the attack, providing insight into how quickly and completely spoofed messages propagated through the autonomy stack. Finally, the Outcome column records the end result on the physical QCar2 platform, serving as the ground-truth measure of operational disruption. These columns are selected to directly map the attack surface to the physical consequences in our testbed. By pairing the spoofed topic with the specific falsification method, then tracking the intermediate control-node reaction and the final physical outcome, we provide a complete trace from cyber-level injection to real-world hazard manifestation.
\vspace{-2mm}

\section{Potential Mitigations}
\vspace{-1mm}
Mitigating the demonstrated vulnerabilities requires measures addressing both the spoofing phase and the initial supply chain compromise. For the latter, hardening the ROS~2 software supply chain is critical. Secure build environments, reproducible package builds, dependency integrity verification, and mandatory code signing reduce the risk of maliciously modified binaries entering deployment. Regular verification of package hashes against trusted registries, combined with isolated build pipelines, further limits the attack surface for code injection. At runtime, topic spoofing risks can be reduced through multi-layer validation. Implementing semantic checks within subscriber nodes ensures received messages fall within physically plausible ranges.

\begin{comment}
    For example, rejecting sudden throttle jumps or improbable distance values can prevent immediate hazardous reactions. Cross-sensor redundancy; such as comparing LiDAR, camera, and odometry readings, can detect injected perception data inconsistent with other modalities. Additionally, rate-limiting critical actuator topics and employing temporal filtering can reduce the impact of high-frequency malicious messages without introducing significant latency. Together, these strategies strengthen resilience against both credential misuse and direct topic injection attacks.
\end{comment}

\vspace{-01mm}
\section{Conclusion}
\vspace{-0.9mm}
This work demonstrated a practical SROS~2 supply chain attack on a real-world autonomous vehicle platform, from credential exfiltration to live control and perception spoofing. To the best of our knowledge, this is the first practical demonstration of an SROS~2 focused supply chain attack, bridging a gap in existing literature that has largely overlooked secure ROS~2 package provenance. Although validated on the QCar2, the approach is directly applicable to any DDS-based robotic platform where SROS~2 keystores form the root of trust. Our results emphasize the need for stronger package verification, keystore protection, and runtime monitoring in secure ROS~2 deployments. Future research will explore lightweight intrusion detection and provenance verification mechanisms to detect malicious package artifacts before deployment.
\vspace{-1mm}

\bibliographystyle{IEEEtran}
\bibliography{reference}

\begin{thebibliography}{10}
\providecommand{\url}[1]{#1}
\csname url@samestyle\endcsname
\providecommand{\newblock}{\relax}
\providecommand{\bibinfo}[2]{#2}
\providecommand{\BIBentrySTDinterwordspacing}{\spaceskip=0pt\relax}
\providecommand{\BIBentryALTinterwordstretchfactor}{4}
\providecommand{\BIBentryALTinterwordspacing}{\spaceskip=\fontdimen2\font plus
\BIBentryALTinterwordstretchfactor\fontdimen3\font minus \fontdimen4\font\relax}
\providecommand{\BIBforeignlanguage}[2]{{%
\expandafter\ifx\csname l@#1\endcsname\relax
\typeout{** WARNING: IEEEtran.bst: No hyphenation pattern has been}%
\typeout{** loaded for the language `#1'. Using the pattern for}%
\typeout{** the default language instead.}%
\else
\language=\csname l@#1\endcsname
\fi
#2}}
\providecommand{\BIBdecl}{\relax}
\BIBdecl

\bibitem{m1_4519604}
E.~A. Lee, ``Cyber physical systems: Design challenges,'' in \emph{2008 11th IEEE International Symposium on Object and Component-Oriented Real-Time Distributed Computing (ISORC)}, 2008, pp. 363--- 369.

\bibitem{m2_7924372}
A.~Humayed, J.~Lin, F.~Li, and B.~Luo, ``Cyber-physical systems security—a survey,'' \emph{IEEE Internet of Things Journal}, vol.~4, no.~6, pp. 1802--- 1831, 2017.

\bibitem{m3_6899663}
J.~Petit and S.~E. Shladover, ``Potential cyberattacks on automated vehicles,'' \emph{IEEE Transactions on Intelligent Transportation Systems}, vol.~16, no.~2, pp. 546--- 556, 2015.

\bibitem{m4_7935369}
J.~Giraldo, E.~Sarkar, A.~A. Cardenas, M.~Maniatakos, and M.~Kantarcioglu, ``Security and privacy in cyber-physical systems: A survey of surveys,'' \emph{IEEE Design \& Test}, vol.~34, no.~4, pp. 7--- 17, 2017.

\bibitem{m5_checkoway2011comprehensive}
S.~Checkoway, D.~McCoy, B.~Kantor, D.~Anderson, H.~Shacham, S.~Savage, K.~Koscher, A.~Czeskis, F.~Roesner, and T.~Kohno, ``Comprehensive experimental analyses of automotive attack surfaces,'' in \emph{20th USENIX security symposium (USENIX Security 11)}, 2011.

\bibitem{m35_11030565}
A.~Solanki, W.~A. Amiri, M.~Mahmoud, B.~Swieder, S.~R. Hasan, and T.~N. Guo, ``Survey of navigational perception sensors' security in autonomous vehicles,'' \emph{IEEE Access}, vol.~13, pp. 104\,937--- 104\,965, 2025.

\bibitem{m6_quigley2009ros}
M.~Quigley, K.~Conley, B.~Gerkey, J.~Faust, T.~Foote, J.~Leibs, R.~Wheeler, A.~Y. Ng \emph{et~al.}, ``Ros: an open-source robot operating system,'' in \emph{ICRA workshop on open source software}, vol.~3, no. 3.2.\hskip 1em plus 0.5em minus 0.4em\relax Kobe, 2009, p.~5.

\bibitem{m7_macenski2022robot}
S.~Macenski, T.~Foote, B.~Gerkey, C.~Lalancette, and W.~Woodall, ``Robot operating system 2: Design, architecture, and uses in the wild,'' \emph{Science robotics}, vol.~7, no.~66, p. eabm6074, 2022.

\bibitem{m10_Fazzari2019}
\BIBentryALTinterwordspacing
K.~Fazzari, ``Ros 2 dds‑security integration,'' Design document, ROS 2 Design (Open Source Robotics Foundation), 2019, date written: July 2019; Last modified: July 2020. [Online]. Available: \url{https://design.ros2.org/articles/ros2_dds_security.html}
\BIBentrySTDinterwordspacing

\bibitem{m11_sros2_repo}
ros2 (Open~Robotics and the ROS 2~community), ``{ros2/sros2}: tools to generate and distribute keys for sros 2,'' \url{https://github.com/ros2/sros2}, 2025, gitHub repository; accessed on 2025‑08‑08.

\bibitem{m31_dosovitskiy2017carla}
A.~Dosovitskiy, G.~Ros, F.~Codevilla, A.~Lopez, and V.~Koltun, ``Carla: An open urban driving simulator,'' in \emph{Conference on robot learning}.\hskip 1em plus 0.5em minus 0.4em\relax PMLR, 2017, pp. 1--- 16.

\bibitem{m30_liu2020novel}
C.~Liu, C.~Zhou, W.~Cao, F.~Li, and P.~Jia, ``A novel design and implementation of autonomous robotic car based on ros in indoor scenario,'' \emph{Robotics}, vol.~9, no.~1, p.~19, 2020.

\bibitem{m8_garzon2017using}
M.~Garz{\'o}n, J.~Valente, J.~J. Rold{\'a}n, D.~Garz{\'o}n-Ramos, J.~de~Le{\'o}n, A.~Barrientos, and J.~del Cerro, ``Using ros in multi-robot systems: Experiences and lessons learned from real-world field tests,'' in \emph{Robot Operating System (ROS) The Complete Reference (Volume 2)}.\hskip 1em plus 0.5em minus 0.4em\relax Springer, 2017, pp. 449--- 483.

\bibitem{m9_dieber2017security}
B.~Dieber, B.~Breiling, S.~Taurer, S.~Kacianka, S.~Rass, and P.~Schartner, ``Security for the robot operating system,'' \emph{Robotics and Autonomous Systems}, vol.~98, pp. 192--- 203, 2017.

\bibitem{m12_deng2022security}
G.~Deng, G.~Xu, Y.~Zhou, T.~Zhang, and Y.~Liu, ``On the (in) security of secure ros2,'' in \emph{Proceedings of the 2022 ACM SIGSAC Conference on Computer and Communications Security}, 2022, pp. 739--- 753.

\bibitem{m13_sros2_issue_282}
amrc benmorrow, ``{````Chain of trust issues with a single CA certificate''},'' \url{https://github.com/ros2/sros2/issues/282}, 2022, gitHub issue \#282, opened on December 13, 2022.

\bibitem{m14_kim2018security}
J.~Kim, J.~M. Smereka, C.~Cheung, S.~Nepal, and M.~Grobler, ``Security and performance considerations in ros 2: A balancing act,'' \emph{arXiv preprint arXiv:1809.09566}, 2018.

\bibitem{m15_portugal2024ros3}
D.~Portugal, R.~P. Rocha, and J.~P. Castilho, ``Inquiring the robot operating system community on the state of adoption of the ros 2 robotics middleware,'' \emph{Int. J. Intell. Robot. Appl.}, vol.~9, pp. 454--- 479, 2025.

\bibitem{m16_yang2024formal}
S.~Yang, J.~Guo, and X.~Rui, ``Formal analysis and detection for ros2 communication security vulnerability,'' \emph{Electronics}, vol.~13, no.~9, p. 1762, 2024.

\bibitem{m17_app15137493}
M.~Sakhai, K.~Sithu, M.~K.~S. Oke, and M.~Wielgosz, ``Cyberattack resilience of autonomous vehicle sensor systems: Evaluating rgb vs. dynamic vision sensors in carla,'' \emph{Applied Sciences}, vol.~15, no.~13, 2025.

\bibitem{m18_jang2014survey}
J.~Jang-Jaccard and S.~Nepal, ``A survey of emerging threats in cybersecurity,'' \emph{Journal of computer and system sciences}, vol.~80, no.~5, pp. 973--- 993, 2014.

\bibitem{m19_kyrofa2019security}
Kyrofa, ``Security issue on ros build farm,'' \url{https://discourse.openrobotics.org/t/security-issue-on-ros-build-farm/9342/2}, May 30 2019, online forum post on Open Robotics Discourse.

\bibitem{m20_ros_buildfarm2025}
R.~I. Team, ``ros\_buildfarm: Ros build farm based on docker,'' \url{https://github.com/ros-infrastructure/ros_buildfarm}, 2025, gitHub repository. Accessed on August 8, 2025.

\bibitem{m33_cisa2020_sunburst}
Cybersecurity and I.~S.~A. (CISA), ``Advanced persistent threat compromise of government agencies, critical infrastructure, and private sector organizations,'' \url{https://www.cisa.gov/news-events/cybersecurity-advisories/aa20-352a}, 2020, alert Code AA20‑352A; last revised April 15, 2021; accessed August 8, 2025.

\bibitem{m21_nist2024_cve2024_3094}
{National Vulnerability Database (NVD), NIST}, ``Cve-2024-3094: Malicious code in upstream xz utils tarballs (impacting liblzma build process),'' \url{https://nvd.nist.gov/vuln/detail/CVE-2024-3094}, 2024, cVSS score 10.0; accessed August 8, 2025.

\bibitem{m22_claburn2024_xz_backdoor}
T.~Claburn, ``Malicious backdoor spotted in linux compression library xz,'' \url{https://www.theregister.com/2024/03/29/malicious_backdoor_xz/}, 2024, the Register — Security section; accessed August 8, 2025.

\bibitem{m23_pypi2024_domain_abuse}
{Python Package Index Blog}, ``Malware distribution and domain abuse,'' \url{https://blog.pypi.org/posts/2024-04-10-domain-abuse/}, 2024, the Python Package Index Blog; accessed August 8, 2025.

\bibitem{m24_toulas2022_pypi_credentials}
B.~Toulas, ``10 malicious pypi packages found stealing developer's credentials,'' \url{https://www.bleepingcomputer.com/news/security/10-malicious-pypi-packages-found-stealing-developers-credentials/}, 2022, bleeping Computer; accessed August 8, 2025.

\bibitem{m25_rti2025_access_control}
{RTI Connext DDS Security}, ``Access control — rti security plugins user's manual,'' \url{https://community.rti.com/static/documentation/connext-dds/current/doc/manuals/connext_dds_secure/users_manual/p2_core/access_control.html}, 2025, accessed August 8, 2025; covers governance and permissions enforcement via Security Plugins.

\bibitem{m26_cova2010detection}
M.~Cova, C.~Kruegel, and G.~Vigna, ``Detection and analysis of drive-by-download attacks and malicious javascript code,'' in \emph{Proceedings of the 19th international conference on World wide web}, 2010, pp. 281--- 290.

\bibitem{m27_yadav2010detecting}
S.~Yadav, A.~K.~K. Reddy, A.~N. Reddy, and S.~Ranjan, ``Detecting algorithmically generated malicious domain names,'' in \emph{Proceedings of the 10th ACM SIGCOMM conference on Internet measurement}, 2010, pp. 48--- 61.

\bibitem{m34_quanser_qcar2}
\BIBentryALTinterwordspacing
{Quanser Inc.}, ``{QCar 2: Sensor‑rich autonomous vehicle for self‑driving applications},'' Product page on Quanser website, 2025, accessed: 2025‑08‑08. [Online]. Available: \url{https://www.quanser.com/products/qcar-2/}
\BIBentrySTDinterwordspacing

\end{thebibliography}

\end{document}